# Expanding the Grammar of Biology

Michel Eduardo Beleza Yamagishi[1*], Roberto Hirochi Herai[2]

[1]Laboratório de Bioinformática Aplicada, Embrapa Informática Agropecuária. Email: michel.yamagishi@embrapa.br.

[2]Department of Cellular & Molecular Medicine, University of California San Diego. Email: rherai@ucsd.edu

*Abstract*

*We metaphorically call "Grammar of Biology" a small field of genomic research, whose main objective is to search for DNA sequence intrinsic properties. Erwin Chargaff inaugurated it back in 50s, but since then little progress has been made. It remained almost neglected until early 90s, when Vinayakumar V. Prabhu made a major contribution determining the Symmetry Principle. Remarkably, during the "Grammar of Biology" development, different sciences have contributed, for instance, Chargaff used his Chemistry background to discover the so-called Chargaff's rules; taking advantage of several publicly available genomic sequences, and through Computational and Statistical analyses, Prabhu identified the Symmetry Principle and, recently, using a Mathematical approach, we have discovered four new Generalized Chargaff's rules. Our work has expanded the "Grammar of Biology", and created the conceptual and theoretical framework necessary to further developments.*

## The Alphabet

Oswald T. Avery (1877–1955) created history in 1944 when he published a memorable article (*1*), which, for the first time, revealed an association between deoxyribonucleic acid (DNA) and the transmission of hereditary characteristics. To measure the importance of his work, we must remember that DNA had been discovered by Johannes Friedrich Miescher (1844–1895) in 1871 (*2*); however, until almost the end of the first half of the $20^{th}$ century, knowledge about this macromolecule and its functions in living organisms was scarce. It was known, for example, that DNA was composed of the nucleotides adenine (A), guanine (G), thymine (T), and cytosine (C); however, for many decades, it was believed that DNA was merely an unvarying and consecutive repetition of these four monomers, i.e., $(AGTC)_n$, where $n$ indicates the number of repeats, which was also completely unknown. This was the so-called tetranucleotide hypothesis, in which DNA was considered as one more polymer among many other. Hence, Avery's work, which reveals DNA's true role, is of outstanding significance.

## The Grammar of Biology

One of the people who quickly realized the potential of Avery's discovery was Erwin Chargaff (1905–2002). In an article commemorating the centenary of the discovery of DNA in 1971, Chargaff recalled that, when reading Avery's work, he saw before him the beginning of a *grammar of biology* (*3*). Moreover, he recalled that the impact on him of this discovery was so significant that he decided to abandon all that he was working on to start from scratch his own research on DNA. The early years were not easy because Chargaff needed to develop more accurate methods for the chemical characterization of nucleic acids. After much work at the bench, he published his first significant results (*4*, *5*), which included Chargaff's first parity rule (CFPR). This rule states that in *double-stranded DNA*, the frequency of adenine is equal to the frequency of thymine, that is, $F(A) = F(T)$, and the frequency of cytosine is equal to the frequency of guanine, that is, $F(C) = F(G)$ (*6*).

CFPR, as summed up in these two equations, seems like a theoretical result without major practical consequences; however, it was actually a magnificent discovery and became one of the clues that, in addition to the X-ray images produced by Rosalind Franklin (1920–1958), resulted in the three-dimensional model of the double-helix structure of DNA (*7*, *8*).

The immediate consequence of the double-helix model is that DNA is formed by two complementary strands. Chargaff soon wondered what would be the individual properties of these strands. He overcame the technical difficulties of his time and estimated the proportions of each nucleotide in *single-stranded DNA* (*9*). He was surprised to find that a weaker version of CFPR, in which the equal sign (=) gives way to the approximately equal sign ($\approx$), was also valid in this case. This is the so-called Chargaff's second parity rule (CSPR), which states that, in *single-stranded* DNA, the frequency of adenine is approximately equal to that of thymine, that is, $F(A) \approx F(T)$, and the frequency of cytosine is approximately equal to that of guanine, that is, $F(C) \approx F(G)$.

With his studies, Chargaff definitively refuted the tetranucleotide hypothesis of the DNA structure, showed that the DNA sequence is more complex than initially assumed, and by defining the rules of parity, opened new horizons for DNA research.

**Expanding the Grammar of Biology**

It was around 2005 that our interest in CSPR started to develop, after empirically rediscovering it in the data from the human genome assembly. Until then, Chargaff was for us, an illustrious unknown. Given our mathematical background, we soon thought about mathematically modeling that phenomenon. Our first work in the area was published after three years (*10*). This study presented a mathematical model that predicted with accuracy the frequencies of nucleotides for each of the chromosomes in the human genome. The main premise was the validity of the CSPR. Although less well known than the first rule, CSPR is, in fact, observed in many organisms (*11*). However, because our model only applied to the human genome sequence and was limited to the frequency of a single nucleotide, there was a significant possibility that our results could have been a statistical artifact, i.e., have resulted from chance. To rule out this possibility, our goals focused on trying to extend the results to other organisms as well as on adapting the model for frequencies of more than one nucleotide, i.e., for oligonucleotides. After several failures, we decided to search the literature for leads that could help us. When reviewing the studies by Donald R. Forsdyke, we found the Bell and Forsdyke conjecture (*12*) according to which CSPR was, in fact, a particular case of a higher-order rule. That attracted our attention because, curiously, in 1993, Vinayakumar V. Prabhu, through the direct calculation of oligonucleotide frequencies from genomic sequences available at the time, had already discovered a generalization of the CSPR, known as the symmetry principle (*13*). This states that, for any given oligonucleotide, its frequency is approximately equal to its complementary nucleotide, that is, $F(w) \approx F(R(C(w)))$, where $R$ is the reverse operator; $C$ the complement operator; and $w$, any oligonucleotide. When applied to a single nucleotide, it is trivial to verify that the symmetry principle is a generalization of the CSPR: taking $w = A$, we have $F(A) \approx F(T)$, and similarly, taking $w = C$, we have $F(C) \approx F(G)$.

However, despite all our efforts, the symmetry principle was not enough for us to achieve our goals. Accordingly, we started to ask ourselves whether there would be any other generalization to the CSPR.

Another clue found in the literature was an article by Chargaff (*14*), in which he described the relationship between the frequencies of nucleotides in double-stranded DNA using four identities, which for the purposes of their adaptation to single-stranded DNA can be rewritten as follows:

$$F(A) + F(G) \approx F(T) + F(C) \qquad (1)$$
$$F(A) \approx F(T) \qquad (2)$$
$$F(C) \approx F(G) \qquad (3)$$
$$F(A) + F(C) \approx F(T) + F(G) \qquad (4)$$

Equations (2) and (3) are the classic representations of the CSPR. It is easy to observe that if (2) and (3) are true, then equations (1) and (4) follow, and vice versa. In other words, we could also enunciate CSPR using identities (1) and (4). Therefore, the following question seemed natural to us: if the symmetry principle can be considered a generalization of CSPR in relation to identities (2) and (3), then what would be the generalization of CSPR in relation to identities (1) and (4)? To address this question, we perceived that we needed to build a conceptual and theoretical framework, because there were questions that could only be answered satisfactorily with clear concepts and an adequate theory. For example, (1) and (4) concern the sum of the frequencies of two nucleotides on each side of the equation; in the case of more than one nucleotide, which and how many oligonucleotides would be on each side of the equation? How many equations would there be? Is there an equation for dinucleotides and one for trinucleotides, and so on?

The answers to all these questions come naturally from the conceptual and theoretical foundation that we published in 2011 (*15*). We will summarize the main results, avoiding the details (*16*), and explain how these were used to answer the questions raised above. The evolution of the conceptual base was slow and incremental, sometimes occurring through the organization of existing concepts, and at other times, by adding new ones. We started from the most basic points, such as an alphabet and a dictionary of words of fixed size, until we reached the formal definition of complementary ($C$) and reverse ($R$) operators, which are used very frequently in the field of bioinformatics. From a theoretical point of view, after exploring and cataloging the main mathematical properties of these operators, our first insight occurred when we realized that operators $C$, $R$, and their respective compositions, called equivalence operators, were defining *equivalence classes* (ECs) in words of a fixed size (dictionary). Namely, operators grouped words into disjoint groups, the union of which was the whole dictionary, that is, the ECs naturally induced a *partition* in the dictionary. This was a fundamental discovery, because suddenly there was one mathematically simple and elegant way to represent the dictionary with a reduced number of words. To formalize this reduced representation of the dictionary, two new concepts were introduced: *generating set* (GS) and *mathematical table* (MT), concepts that we have attempted to explain in the next two paragraphs.

ECs, as the name suggests, are sets in which there is equivalence between elements; consequently, each EC can be unequivocally represented by any one of its elements. Thus the GS is the set formed by a representative of each EC. From the elements of the GS, and using the equivalence operators, it is possible to obtain all the other elements of each EC. The concept is simple, but powerful. Note that, when words have a fixed size equal to three letters, the dictionary has 64 words, but the GS has only 20. Thus, with only 20 words, it is possible to represent the entire dictionary of 64 words, grouped in their respective ECs. It is important to highlight that the GS definition does not include any criteria for the choice of representatives from each EC, which not only implies that GS is not unique but also suggests that, from the mathematical point of view, the choice of GS elements is arbitrary. Obviously, this does not exclude the existence of selection criteria of other nature.

MT, in turn, is simply the matrix representation of the EC, in which the number of rows in the matrix corresponds to the number of elements in the GS, and the number of columns is always equal to 4, with the first column formed by elements of GS and the other three formed by the complementary reverse, complementary, and reverse of their GS elements, respectively (see

Table 1). We chose the name MT to emphasize the mathematical nature of the table. It is not similar to the 'Codon table', which, for example, is associated with a biological phenomenon.

By arranging the words in the MT, we discovered a pattern that, if it had not been for the MT, would have remained hidden; this was our second insight, as it was observed, regardless of the choice of GS:

$$\sum_{i=1}^{t} F(g_i) \approx \sum_{i=1}^{t} F(C(g_i)), \quad (5)$$

$$\sum_{i=1}^{t} F(R(g_i)) \approx \sum_{i=1}^{t} F(R(C(g_i))), \quad (6)$$

where $g_i \in GS$, $1 \leq i \leq t$, and $t$ is the number of GS elements.

Although it might not be obvious at first sight, equations (5) and (6) are the desired generalizations of (1) and (4). To demonstrate this, consider the simplest case of a single nucleotide where the alphabet is equal to the dictionary; that is, $\{A,C,G,T\}$. Using equivalence operators, we have only two EC: $\{A,T\}$ and $\{C,G\}$. We can choose the set $\{A,C\}$ as the GS. Thus, $t=2$ and $g_1 = A$ e $g_2 = C$. Because the operator $R$ plays no role in this case, equations (5) and (6), for that choice of GS, would be equal to $F(A)+F(C) \approx F(T)+F(G)$; that is, equation (4); similarly, if we choose GS as $\{A,G\}$, we would have $F(A)+F(G) \approx F(T)+F(C)$; that is, equation (1). Therefore, it is proved that we obtained the alternative generalization to CSPR by using the concepts of GS and MT.

With this new formalization, we can return with more confidence to the questions raised before and provide the following answers:

(i) Which oligonucleotides do we add on each side of the equation? In equation (5), we will add up the frequencies of the oligonucleotides of GS on one side of the equation and the frequencies of the complementary forms for the same elements of GS on the other side; similarly, in equation (6), we will add up frequencies of the reverse forms on one side, and the frequencies of the reverse complementary forms of the GS elements on the other side.

(ii) How many oligonucleotides will we add up? It depends on the number of elements in GS. For words with a single nucleotide, we will have $t=2$; for two nucleotides, $t=6$; for three nucleotides, $t=20$, and in general, for words with $k$ nucleotides, $t = 2^{k-1} + 4^{k-1}$

(iii) How many equations would there be? Would there be an equation for dinucleotides and one for trinucleotides, and so on? In its most general form, there would be only two equations, and they do not depend on the number of nucleotides in each word. That is, the equations have the same form and are self-similar, independent of the number of nucleotides.

Finally, by using the definition of frequency, we know that:

$$F(A)+F(T)+F(C)+F(G) = 1,$$

and assuming the validity of CSPR, it is possible to show that:

$$F(A) + F(C) \approx \frac{1}{2}, \qquad (a)$$

$$F(T) + F(G) \approx \frac{1}{2}. \qquad (b)$$

Given that (5) and (6) are generalizations of the CSPR, would there also be generalizations to (a) and (b)? The answer is yes, and this was our third *insight*. Proceeding analogously to our work from 2008, taking into account that, by definition, the sum of frequencies of all words, regardless of the number of nucleotides, is always equal to one; and adding that information to equations (5) and (6), it is possible to demonstrate that:

$$\sum_{i=1}^{t} F(g_i) + \sum_{i=1}^{t} F(R(g_i)) \approx \frac{1}{2}, \qquad (7)$$

$$\sum_{i=1}^{t} F(C(g_i)) + \sum_{i=1}^{t} F(C(R(g_i))) \approx \frac{1}{2}. \qquad (8)$$

Equations (7) and (8) are the generalizations of (a) and (b). Thus, it becomes necessary to know how they would behave when applied to real data; their importance could end up just being theoretical if they fail. To assess its practical value, we selected 36 genomic sequences from NCBI (*17*) and calculated equation (7) for all of them by using varying numbers of nucleotides $k$.

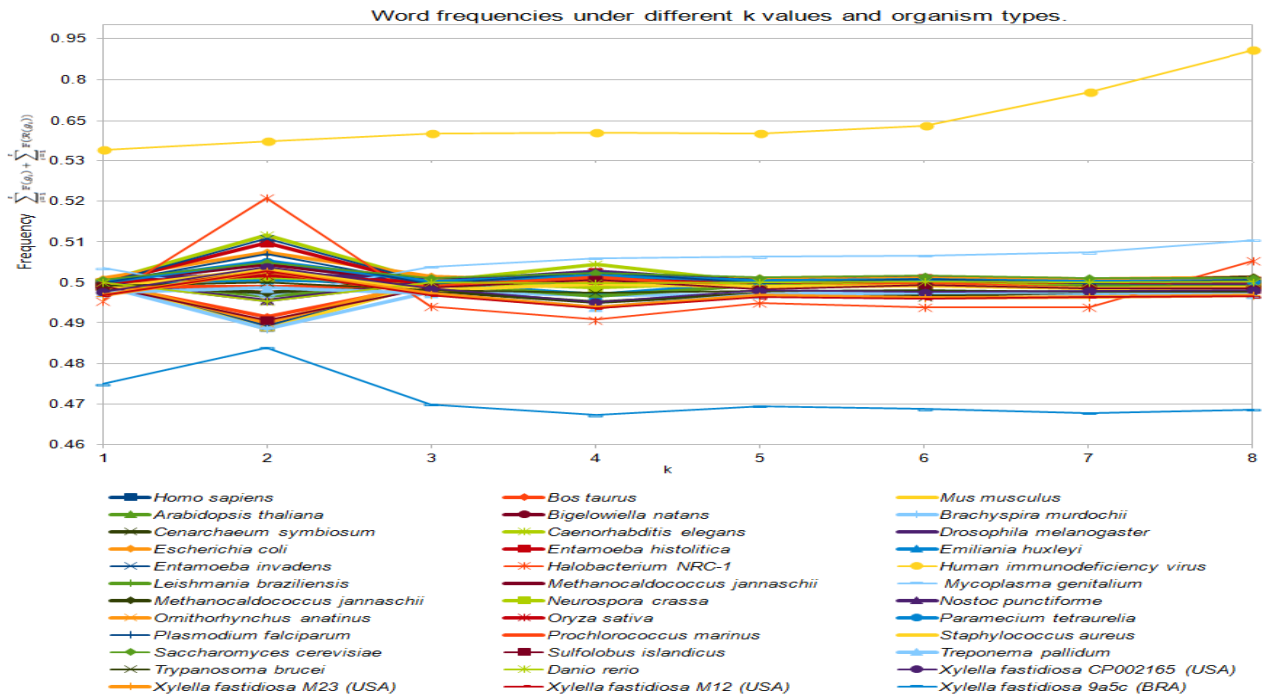

**Fig. 1.** Equation 7 applied to several genomic sequences.

Figure 1 shows that only two genomic sequences are located away from the 0.5 value predicted by equations (7) and (8). The two exceptions are the HIV virus sequence and the assembled sequence for the bacterium *Xylella fastidiosa 9a5c*; nevertheless, two other assembled sequences for *Xylella fastidiosa* satisfied equations (7) and (8). It is interesting to note that the other 34 tested sequences seem to use the same grammar. Among them, sequences of bacteria, fungi, plants, birds, fish, and mammals indicated that the results could be extended to a diverse range of species by simply expanding the *grammar of biology* as introduced by Chargaff.

In reflecting on the discovery of the double-helix structure of DNA, Chargaff stated that *"if Rosalind Franklin and I could have collaborated, we might have come up with something of the sort in one or two years. I doubt, however, that we could ever have elevated the double helix into the mighty symbol that has replaced the cross as the signature of the biological alphabet."* Statements like to this led some people to consider Chargaff as a resentful person. In our opinion, this statement demonstrates how he was misunderstood. Chargaff had a classical erudition that made him a *sui generis* scientist, as can be observed in his memoirs (*18*). Influenced by Karl Kraus (1874–1936), Chargaff was often ironic (*19*), which earned him some disaffected relationships. However, even at the risk of incomprehension, Chargaff did not hesitate to express deep thoughts through simple sentences, as in the following example: *"We posit intelligence where we deny it. We humanize things, but we reify man."*

**Acknowledgements:** We dedicate this study to the memories of the great scientist Erwin Chargaff, and to Marlene Maria Fontes Belleza (first author's mother). We are thankful to CNPq for financially supporting this research and to colleagues from the Multiuser Laboratory of Bioinformatics at Embrapa for their encouragement and support.

**Table 1: Mathematical table for codes consisting of three nucleotides each**

| Class | $g_i$ | $C(R(g_i))$ | $C(g_i)$ | $R(g_i)$ |
|---|---|---|---|---|
| 1 | AAA | - | TTT | - |
| 2 | AAT | ATT | TTA | TAA |
| 3 | TTG | CAA | AAC | GTT |
| 4 | CTT | AAG | GAA | TTC |
| 5 | ATA | - | TAT | - |
| 6 | ATC | GAT | TAG | CTA |
| 7 | ATG | CAT | TAC | GTA |
| 8 | ACA | - | TGT | - |
| 9 | TGA | TCA | ACT | AGT |
| 10 | CCA | TGG | GGT | ACC |
| 11 | GCA | TGC | CGT | ACG |
| 12 | TCT | - | AGA | - |
| 13 | GCT | AGC | CGA | TCG |
| 14 | AGG | CCT | TCC | GGA |
| 15 | CAC | - | GTG | - |
| 16 | CAG | CTG | GTC | GAC |
| 17 | CTC | - | GAG | - |
| 18 | CCC | - | GGG | - |
| 19 | GCC | GGC | CGG | CCG |
| 20 | GCG | - | CGC | - |